\documentclass[lettersize,journal]{IEEEtran}
\usepackage{amsmath,amsfonts}
\usepackage{algorithmic}
\usepackage{array}
\usepackage[caption=false,font=normalsize,labelfont=sf,textfont=sf]{subfig}
\usepackage{textcomp}
\usepackage{stfloats}
\usepackage{xcolor}
\usepackage{url}
\usepackage{cite}
\usepackage{hyperref}
\usepackage{verbatim}
\usepackage{graphicx}
\usepackage{amsfonts}
\usepackage{soul}
\def\BibTeX{{\rm B\kern-.05em{\sc i\kern-.025em b}\kern-.08em
    T\kern-.1667em\lower.7ex\hbox{E}\kern-.125emX}}
\usepackage{balance}
\hypersetup{
     colorlinks   = true,
     linkcolor    = blue,
     citecolor    = blue,
     urlcolor     = blue,
     filecolor    = blue,
     pdftitle={Brain-inspired Beam Selection for mmWave RIS Communication}
}
\begin{document}
\title{Brain-inspired Beam Selection for mmWave RIS Communication}
\author{Aris Karampelas Timotijevic, 
Christos G. Tsinos,~\IEEEmembership{Senior Member,~IEEE},\\
Alexandros-Apostolos A. Boulogeorgos,~\IEEEmembership{Senior Member,~IEEE}, and Theodoros A. Tsiftsis,~\IEEEmembership{Senior Member,~IEEE}
\thanks{The work of A. Karampelas Timotijevic, T. A. Tsiftsis, and C. G. Tsinos was supported by the research project NEURONAS. The research project NEURONAS is implemented in the framework of H.F.R.I. call ``3rd Call for H.F.R.I.'s Research Projects to Support Faculty Members \& Researchers" (H.F.R.I. Project Number: 25726).}
\thanks{A. Karampelas Timotijevic and T. A. Tsiftsis are with the Department of Informatics and Telecommunications, University of Thessaly, Lamia 35100, Greece (e-mail: \{akarampel;  tsiftsis\}@uth.gr).}
\thanks{C. G. Tsinos is with the Department of Informatics \& Telecommunications, National and Kapodistrian University of Athens, Athens (e-mail:  ctsinos@di.uoa.gr).}
\thanks{A. -A. A. Boulogeorgos is with the Department of Electrical and Computer Engineering, Democritus University of Thrace, Xanthi, Greece (e-mail: abouloge@ee.duth.gr).}}


\maketitle

\begin{abstract}
Beam selection at large passive reconfigurable intelligent surfaces (RISs) requires either full channel knowledge or beam training, both of which entail unwanted overhead. On the other hand, learned models that predict the optimal codeword from channels sampled at a few active elements avoid both, at the cost of energy-demanding multiply-and-accumulate operations. In this paper, we propose spiking neural network (SNN)-based beam selection for a semi-active RIS assisting a mmWave multiple-input single-output communication system. Direct and sigmoid rate coding are compared against a multi-layer perceptron (MLP) of identical architecture. Energy analysis further reveals that the input layer dominates SNN energy consumption.
\end{abstract}

\begin{IEEEkeywords}
Beam selection, mmWave, reconfigurable intelligent surfaces (RISs), spiking neural networks (SNNs).
\end{IEEEkeywords}

\section{Introduction}

Reconfigurable intelligent surface (RIS) is a key enabler for high-frequency 6G communications, restoring connectivity when the direct path is obstructed \cite{Boulogeorgos2020}. At millimeter-wave (mmWave) frequencies, the reflected link suffers a double path loss, so the surface needs thousands of meta-elements to provide a useful gain \cite{Taha2021}. Selecting the reflection beam that maximizes the rate at the user then becomes costly. Estimating the channel at every element requires an RF chain per element, i.e., a fully active surface, whereas beam training sweeps a codebook whose size grows with the number of elements, at a prohibitive overhead \cite{Wei2022}.

To address this, deep learning (DL) has been proposed to avoid both requirements. For example, Taha \textit{et al.} in \cite{Taha2021} predicted the optimal codeword from channels sampled at a few active elements. The DL model, however, runs on the RIS controller, whose energy supply is limited, and every inference costs millions of multiply-and-accumulate (MAC) operations. Spiking neural networks (SNNs) target this cost, replacing MACs with accumulations (ACs) triggered only by binary spikes, which neuromorphic processors execute at low energy \cite{Davies2018}. SNNs have been applied to RIS phase-shift optimization \cite{Tsinos2024}, to beamforming in vehicular integrated sensing and communication \cite{Shang2026}, and to RIS beam training with spiking input encoders \cite{Urakami2023}. These works, however, do not examine how the energy per inference of the SNN depends on the system parameters or on the spike encoding of its input.

To bridge the aforementioned critical gap, and in sharp contrast to the energy-intensive conventional DL model of \cite{Taha2021} and the unsupervised continuous-phase design of \cite{Tsinos2024}, our technical contribution is as follows:
\begin{itemize}
    \item We formulate a codebook-constrained reflection beam selection for a semi-active RIS-assisted mmWave multiple-input single-output (MISO) orthogonal frequency division multiplexing (OFDM) system as a supervised regression problem.
    \item We design an SNN that predicts each codeword's maximum achievable rate from the respective channel coefficients sampled at a few randomly placed active elements. Hence, neither full channel state information nor exhaustive beam training is needed.
    \item We compare direct coding and sigmoid rate coding to an MLP with identical architecture. Moreover, the rate-energy trade-off is characterized as a function of the number of active elements.
\end{itemize}

\textbf{Notation}: $\mathbf{X}, \mathbf{x}, x$, and $\mathcal{X}$ denote a matrix, vector, scalar, and set, with $\vert\mathcal{X}\vert$ the cardinality of $\mathcal{X}$, $\vert x\vert$ the modulus of $x$, $x_i$ the $i$-th entry of $\mathbf{x}$, and $X_{i,j}$ the $\left(i,j\right)$-th entry of $\mathbf{X}$. $\left(\cdot\right)^\mathsf{T}$, $\left(\cdot\right)^\mathsf{H}$, $\|\cdot\|$, $\mathrm{vec}\left(\cdot\right)$, and $\mathrm{diag}\left(\mathbf{x}\right)$ denote transpose, conjugate transpose, Euclidean norm, column-wise vectorization, and the diagonal matrix formed from $\mathbf{x}$, while $\Re\{\cdot\}$ and $\Im\{\cdot\}$ denote the real and imaginary parts. $\mathrm{Bernoulli}\left(p\right)$ and $\mathcal{CN}\left(\mu,{\sigma}^2\right)$ denote the Bernoulli distribution with success probability $p$ and the complex Gaussian distribution. $\sigma\left(\cdot\right)$ is the sigmoid function.

\section{System Model}

In this paper, we consider an RIS-assisted MISO downlink system in the mmWave band. The RIS is deployed at a fixed location to assist the data transmission from the base station (BS) to a single-antenna user. The location of the RIS is known a priori to the BS. In this scenario, the direct link between the BS and the user is blocked and the RIS is placed to provide a virtual line-of-sight (LoS) link. The BS is equipped with $N$ antennas, while the RIS employs a total of $M$ elements. We also adopt an OFDM-based transmission with $K$ subcarriers. Let $\mathbf{G}_k\in\mathbb{C}^{M\times N}$ be the channel between the BS and the RIS, and $\mathbf{h}_k\in\mathbb{C}^{M}$ be the channel between the RIS and the user, at the $k$-th subcarrier, respectively. The received signal at the user is expressed as
\begin{equation}\label{eq:signal}
    y_k=\mathbf{h}_k^\mathsf{T}\mathbf{\Psi}\mathbf{G}_k\mathbf{w}_kx_k+n_k,
\end{equation}
where $x_k$ is the transmitted symbol, $\mathbf{\Psi}=\mathrm{diag}\left(\boldsymbol{\psi}\right)\in\mathbb{C}^{M\times M}$ is the RIS reflection matrix, with $\boldsymbol{\psi}\in\mathbb{C}^{M}$ the reflection vector, and $\mathbf{w}_k\in\mathbb{C}^N$ denotes the beamforming vector used by the BS at the $k$-th subcarrier. The receiver noise is denoted by $n_k \sim \mathcal{CN}\left(0,\sigma_n^2\right)$. Note that, in the above model, paths that reach the RIS more than once are ignored, as their additional hops render their power negligible.

We also assume that the channel knowledge is not available at the RIS. To counterbalance this, the RIS employs a semi-active operation regime \cite{Taha2021}. The operation of the RIS occurs in two phases. First, both the BS and the user send pilot symbols to the RIS. After estimating both channels the RIS controller initiates the beam reflection. The channel is reconstructed from independent samples acquired by randomly scattered active elements on the RIS. Each active element is connected to an RF chain for channel estimation. The channels enter (\ref{eq:signal}) only through the cascaded BS--RIS--user channel $\mathbf{A}_k=\mathrm{diag}\left(\mathbf{h}_k\right)\mathbf{G}_k\in\mathbb{C}^{M\times N}$, as $\mathbf{h}_k^\mathsf{T}\mathbf{\Psi}\mathbf{G}_k=\boldsymbol{\psi}^\mathsf{T}\mathbf{A}_k$. For a given $\boldsymbol{\psi}$ and $\mathbf{w}_k$, the effective channel coefficient is therefore
\begin{equation}\label{eq:eff_chan}
    h_{\mathrm{eff},k} = \boldsymbol{\psi}^\mathsf{T}\mathbf{A}_k\mathbf{w}_k.
\end{equation}

The sampled channel, $\mathbf{\bar{H}}_k\in\mathbb{C}^{M_a\times N}$, at the RIS is obtained by acquiring the channel estimates at $M_a$ sparsely placed active elements, for each of the $N$ BS antennas, and for $K_\mathrm{DL}$ of the $K$ subcarriers, as
\begin{equation}\label{eq:sampled_chan}
    \mathbf{\bar{H}}_k = \mathbf{S}\,\mathrm{diag}\left(\mathbf{h}_k+\mathbf{n}_{1,k}\right)\left(\mathbf{G}_k+\mathbf{N}_{2,k}\right),\quad k=1,\ldots,K_\mathrm{DL},
\end{equation}
where $\mathbf{S}\in\{0,~1\}^{M_a\times M}$ is a sampling matrix that selects the active elements, and $\mathbf{n}_{1,k}\in\mathbb{C}^{M}$ and $\mathbf{N}_{2,k}\in\mathbb{C}^{M\times N}$ are the channel-estimation noise vector and matrix, whose entries are independent $\mathcal{CN}\left(0,\sigma_e^2\right)$ across elements, antennas and subcarriers.

To achieve reduced energy consumption at the RIS, each meta-element of the surface is implemented with a discrete phase shifter, which supports a quantized set of angles. To capture this constraint, we assume that the reflection beamforming vector $\boldsymbol{\psi}$ can only be picked from a pre-defined codebook $\mathcal{C}$. Every candidate reflection beamforming codeword $\boldsymbol{\psi}_c$ in $\mathcal{C}$ is assumed to be implemented using discrete phase shifters. In conventional far-field RIS beam training based on angular steering codebooks, the reflection codebook size is typically chosen on the order of the number of RIS elements, i.e., $\left\vert\mathcal{C}\right\vert=M$, so as to sample the angular space with array-limited resolution \cite{Taha2021, Wei2022}.

Given (\ref{eq:eff_chan}), our objective is to design the RIS reflection beamforming vector, $\boldsymbol{\psi}$, and beamforming vector, $\mathbf{w}_k$, to maximize the achievable rate at the receiver, which can be formulated as
\begin{equation}\label{eq:rate}
   R = \frac{1}{K}\sum_{k=1}^{K}\log_2\left(1+\frac{1}{\sigma_n^2}\left\vert\boldsymbol{\psi}^\mathsf{T}\mathbf{A}_k\mathbf{w}_k\right\vert^2\right).
\end{equation}
Each subcarrier has the same power under an equal power allocation scheme, with $P_t$ being the total transmission power. The optimization problem can be formulated as
\begin{equation}\label{eq:optimization}
    \max_{\mathbf{w}_k,\boldsymbol{\psi}}\; R \quad \text{s.t.} \quad \boldsymbol{\psi} \in \mathcal{C},\quad \|\mathbf{w}_k\|^2 = P_t/K.
\end{equation}

The optimization problem in (\ref{eq:optimization}) has no closed-form solution, and $\boldsymbol{\psi}$ and $\mathbf{w}_k$ are coupled. We decouple them with maximal ratio transmission (MRT) at the BS, whose optimal beamforming vector is \cite{Wang2020, Cheng2022}
\begin{equation}\label{eq:mrt}
    \mathbf{w}_k^\star = \sqrt{\frac{P_t}{K}}\,\frac{\left(\boldsymbol{\psi}^\mathsf{T}\mathbf{A}_k\right)^\mathsf{H}}{\left\|\boldsymbol{\psi}^\mathsf{T}\mathbf{A}_k\right\|}.
\end{equation}

By applying (\ref{eq:mrt}), (\ref{eq:optimization}) reduces to an exhaustive search over the codebook,
\begin{equation}
    \begin{aligned}
    \begin{aligned}
    c^{\star} = \; & \arg{\max_{c}{\;}}\frac{1}{K}\sum_{k=1}^{K}\log_2\left(1+\frac{P_t}{K\sigma_n^2}\left\|\boldsymbol{\psi}_c^\mathsf{T}\mathbf{A}_k\right\|^{2}\right)\\
    &\text{s.t.}\quad c\in\left\{1,\ldots,\left|\mathcal{C}\right|\right\},\\
    & \quad \quad{\boldsymbol{\psi}}_{c}\in\mathcal{C}
    \end{aligned}
    \end{aligned}
\end{equation}
This search, however, requires full channel knowledge, i.e., a fully active RIS, which is prohibitive in hardware cost and energy, while beam training incurs an excessive overhead even under the codebook constraint. Therefore, a learned model instead interpolates the missing channel information from the sampled channels in (\ref{eq:sampled_chan}), and an SNN further improves its energy efficiency.

 
\section{Proposed Spiking Architecture}\label{section:3}

\subsection{Input Features}

To train the learned models, the sampled channels of each measurement are stacked over the $K_\mathrm{DL}$ subcarriers into $\bar{\mathbf{h}}=\mathrm{vec}\left(\left[\bar{\mathbf{H}}_1,\ldots,\bar{\mathbf{H}}_{K_\mathrm{DL}}\right]\right)\in\mathbb{C}^{K_\mathrm{DL}M_aN}$, whose real and imaginary parts form the input feature vector
\begin{equation}
    \mathbf{z}=\left[\Re\{\bar{\mathbf{h}}\}^\mathsf{T},~\Im\{\bar{\mathbf{h}}\}^\mathsf{T}\right]^\mathsf{T}\in\mathbb{R}^{2K_\mathrm{DL}M_aN}.
\end{equation}
The features are normalized by the largest channel modulus in the dataset, i.e.,
\begin{equation}\label{eq:maxnorm}
    \tilde{\mathbf{z}} = \frac{\mathbf{z}}{\Delta}, \qquad \Delta = \max_{u\in\mathcal{U},\,i}\left\vert\bar{h}_i^{(u)}\right\vert,
\end{equation}
where $\bar{\mathbf{h}}^{(u)}$ is the stacked vector of measurement $u$ and $\mathcal{U}$ is the set of measurements, so that $\tilde{\mathbf{z}}\in\left[-1,~1\right]^{2K_\mathrm{DL}M_aN}$.
\subsection{Input Encoding}

Two common encoding schemes for producing spike inputs are direct coding and rate coding \cite{Eshraghian2023}. Each encoding scheme leads to different assumptions about hardware implementation and energy consumption of the SNN \cite{Kim2022}.

Under direct coding, the features drive the first layer directly, i.e., $\mathbf{p}=\tilde{\mathbf{z}}$, as in conventional DL. The first layer thus acts as the spike encoder for the rest of the network, and its operations are MACs.

Under rate coding, each feature $p_i$ is converted into a spike train $\mathbf{s}_i=[s_{i,1},\ldots,s_{i,T}]^\mathsf{T}$ of $T$ independent Bernoulli trials, $s_{i,t}\sim \mathrm{Bernoulli}\left(p_i\right)$. As the success probabilities must lie in $\left[0,~1\right]$, the signed features are mapped as $\mathbf{p}^\mathrm{sig}=\sigma\left(-\rho\tilde{\mathbf{z}}\right)$, with $\rho=G\Delta$, so that $\rho\tilde{\mathbf{z}}=G\mathbf{z}$, i.e., the fixed gain $G$ rescales the unnormalized channel coefficients, which are small due to the double path loss, to the sensitive range of the sigmoid.

\subsection{SNN Architecture}\label{section:architecture}

In this work, we employ the leaky integrate-and-fire (LIF) neuron model with reset to zero \cite{Eshraghian2023}, whose spike generation is described by
\begin{equation}\label{eq:lifspike}
    s_{t,i}^{(l)} = \Theta\!\left(u_{t,i}^{(l)} - u_{\mathrm{th},i}^{(l)}\right),
\end{equation}
where $u_{t,i}^{(l)}$ is the membrane potential at time step $t$, for the $i$-th neuron of the $l$-th layer of the SNN, $u_{\mathrm{th},i}^{(l)}$ is the threshold potential at the $i$-th neuron of the $l$-th layer, and $\Theta\left(\cdot\right)$ is the Heaviside step function. For the $0$-th time step we have $u_{0,i}^{(l)} = 0$ and $s_{0,i}^{(l)} = 0$. The corresponding discrete-time membrane dynamics are
\begin{equation}\label{eq:lifneuron}
    u_{t,i}^{(l)} = \beta\left(1-s_{t-1,i}^{(l)}\right) u_{t-1,i}^{(l)} + \sum_j{W_{i,j}^{(l)}s_{t,j}^{(l-1)}},
\end{equation}
where $W_{i,j}^{(l)}$ is the weight connecting pre-synaptic neuron $j$ to
post-synaptic neuron $i$, $\beta$ is the membrane-potential decay rate. When a neuron spikes, its accumulated potential is discarded, so at the next time step it integrates only the fresh input current.

In regression tasks, the output layer typically consists of leaky-integrator (LI) neurons to accumulate incoming spikes without generating output spikes \cite{Rancon2022}. This is achieved by disabling the reset mechanism and setting the firing threshold sufficiently high to prevent spike generation. The LI output therefore provides a continuous-valued representation through membrane readout, which is appropriate for the regression task at hand \cite{Henkes2024}.

\subsection{Computational Energy Model}

Inference cost is measured in synaptic operations. A dense layer performs $N_\mathrm{in}^{(l)}N_\mathrm{out}^{(l)}$ MAC operations, so a conventional MLP regressor with $N_l$ layers requires
\begin{equation}\label{eq:mac_mlp}
    O_\mathrm{MLP} = \sum_{l=1}^{N_l}{N_\mathrm{in}^{(l)}N_\mathrm{out}^{(l)}},
\end{equation}
operations per inference. In a spiking layer, by contrast, an operation is triggered only when a presynaptic neuron fires, and each such event costs a single accumulation (AC), since the binary spikes reduce the product $W_{i,j}^{(l)}s_{t,j}^{(l-1)}$ in (\ref{eq:lifneuron}) to a conditional addition of the weight. The relevant quantity is therefore the spike count per neuron per inference,
\begin{equation}\label{eq:zeta}
    \zeta^{(l)} = \frac{O_\mathrm{spike}^{(l)}}{O_\mathrm{neuron}^{(l)}},
\end{equation}
where $O_\mathrm{spike}^{(l)}$ counts the spikes emitted across all $T$ time steps by the $O_\mathrm{neuron}^{(l)}=N_\mathrm{in}^{(l)}$ presynaptic neurons of layer $l$, i.e., by the input spike trains for $l=1$, so that $\zeta^{(l)}$ is bounded by $T$ rather than by unity. A rate-coded SNN then requires
\begin{equation}\label{eq:rc_snn_flops}
    O_\mathrm{SNN}^\mathrm{rate} = \sum_{l=1}^{N_l}{\zeta^{(l)}N_\mathrm{in}^{(l)}N_{\mathrm{out}}^{(l)}}
\end{equation}
accumulations \cite{Kundu2021}.

Under direct coding, the first layer receives continuous values, so its operations are MACs. The input current is constant over the simulation window and is evaluated once per inference, giving
\begin{equation}\label{eq:dc_snn_flops}
    O_\mathrm{SNN}^\mathrm{drct} = N_\mathrm{in}^{(1)}N_\mathrm{out}^{(1)} + \sum_{l=2}^{N_l}{\zeta^{(l)}N_\mathrm{in}^{(l)}N_\mathrm{out}^{(l)}},
\end{equation}
where the first spiking layer acts as the encoder for the remainder of the network. The energy per inference is
\begin{equation}\label{eq:consumption}
    E = E_{\mathrm{MAC}}O_{\mathrm{MAC}} + E_{\mathrm{AC}}O_{\mathrm{AC}},
\end{equation}
with $O_\mathrm{MAC} = N_\mathrm{in}^{(1)}N_\mathrm{out}^{(1)}$ and $O_\mathrm{AC}$ the summation term of (\ref{eq:dc_snn_flops}) for direct coding, $O_\mathrm{MAC}=0$ and $O_\mathrm{AC}=O_\mathrm{SNN}^\mathrm{rate}$ for rate coding, and $O_\mathrm{MAC} = O_\mathrm{MLP}$, $O_\mathrm{AC}=0$ for the MLP. Following \cite{Kundu2021}, (\ref{eq:consumption}) accounts for synaptic operations, which dominate the inference cost of fully connected architectures of this size. A spiking layer is cheaper than its MAC counterpart only while $\zeta^{(l)} < E_\mathrm{MAC}/E_\mathrm{AC} \approx 5.1$. The final $\arg\max$ over the $|\mathcal{C}|$ outputs adds $|\mathcal{C}|$ comparisons, which is negligible.

\section{Simulation and Results}\label{section:4}
\subsection{Simulation setup}

In this section, the SNN-based beam selection is evaluated against the MLP in terms of achievable rate and inference energy. Channels are generated with DeepMIMO \cite{DeepMIMO2019} for the O1 scenario at $28~\mathrm{GHz}$, with a single path ($L=1$) per link, a bandwidth of $B=100~\mathrm{MHz}$ and $K=512$ subcarriers, of which the first $K_\mathrm{DL}=64$ are used as learned model inputs. The BS employs a $2\times2$ uniform planar array ($N=4$) with transmit power $P_t=5~\mathrm{dBW}$ at a fixed position (row $850$, column $90$ of the O1 user grid), and the RIS has $64\times64$ ($M=4096$) elements with half-wavelength spacing. 

The reflection codebook is a DFT codebook with $|\mathcal{C}|=M$ codewords, and $M_a\in\{4,8,16\}$ active elements are selected at random, with $M_a=8$ unless stated otherwise. The $36200$ candidate receiver positions (rows $1000$ to $1199$) are split at random into $30000$ training and $6200$ evaluation users, so training and evaluation positions are spatially interleaved. Each user corresponds to a receiver position, i.e., a distinct channel realization and optimal codeword, so that the training size is the number of positions observed by the models. 

The sampled measurements at the $M_a$ active elements are corrupted by channel-estimation noise, whereas the ``ground-truth" labels are computed from the noiseless cascaded channel. The channel-estimation noise variance, $\sigma_e^2$, is the noise power per subcarrier, $-204~\mathrm{dBW/Hz}+10\log_{10}(B/K)$ with a $5~\mathrm{dB}$ noise figure and a $10~\mathrm{dB}$ processing gain, i.e., $-156.1~\mathrm{dBW}$, normalized by the $5~\mathrm{dBW}$ transmit power and the $3~\mathrm{dBi}$ antenna gains at the BS and the user, i.e., $\sigma_e^2\approx-167.1~\mathrm{dB}$.

All models minimize the mean square error to the training targets, i.e., the achievable rates of the $|\mathcal{C}|$ codewords, normalized per user by their maximum so that the model is not biased towards users with strong channels. Moreover, all models are trained with the AdamW optimizer, i.e., Adam with decoupled weight decay, with momentum parameters $(0.9,~0.999)$, a batch size of $500$ and no dropout, for at most $200$ epochs, stopping after $8$ epochs without improvement of the validation loss and retaining the best weights. The MLP uses a learning rate of $10^{-3}$, halved every $20$ epochs, and a weight decay of $10^{-2}$, whereas the SNNs use a constant learning rate of $10^{-4}$ without weight decay. The SNNs are trained by backpropagation through time with $T=5$, a fast-sigmoid surrogate gradient of slope $25$, with the LIF thresholds initialized uniformly in $\left[0,~1\right]$, a membrane decay rate $\beta=0.9375=1-2^{-4}$, which allows the leak to be computed with a bit shift and a subtraction in fixed-point neuromorphic hardware, and a reset to zero, while the output LI neurons use $\beta=1$ and no reset, and the rate encoder uses $G=10^{12}$, i.e., $\rho\approx8.3$ for $M_a=8$.

To remain deployable on neuromorphic processors, whose neurons and synapses share a limited on-chip memory \cite{Davies2018}, the network size is kept bounded. Hidden widths that scale with the codebook, e.g., $|\mathcal{C}|$, $4|\mathcal{C}|$ and $4|\mathcal{C}|$, would require about $4.2\times10^{8}$ weights for a $64\times64$ RIS with $M_a=8$. All models therefore share $N_l=4$ weight layers, with $2K_\mathrm{DL}M_aN$ inputs, three hidden layers of $500$ neurons, selected empirically, and $|\mathcal{C}|$ outputs, i.e., about $4.6\times10^{6}$ weights for $M_a=8$. In contrast to $\beta$, the thresholds are optimized jointly with the weights, so that each layer adapts its firing sparsity to the dynamic range of its input current.

\subsection{Achievable Rate}

\begin{figure}[!t]
\centering
\captionsetup[subfloat]{font=footnotesize}
\subfloat[]{%
  \includegraphics[width=0.47\columnwidth]{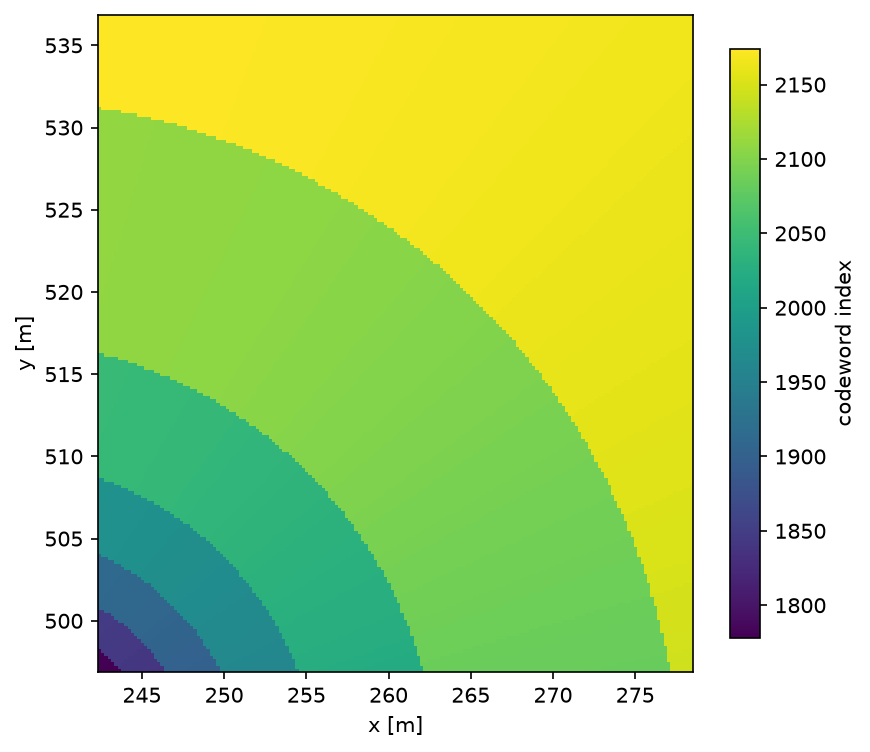}%
  \label{fig:beam_opt}}
\hfil
\subfloat[]{%
  \includegraphics[width=0.47\columnwidth]{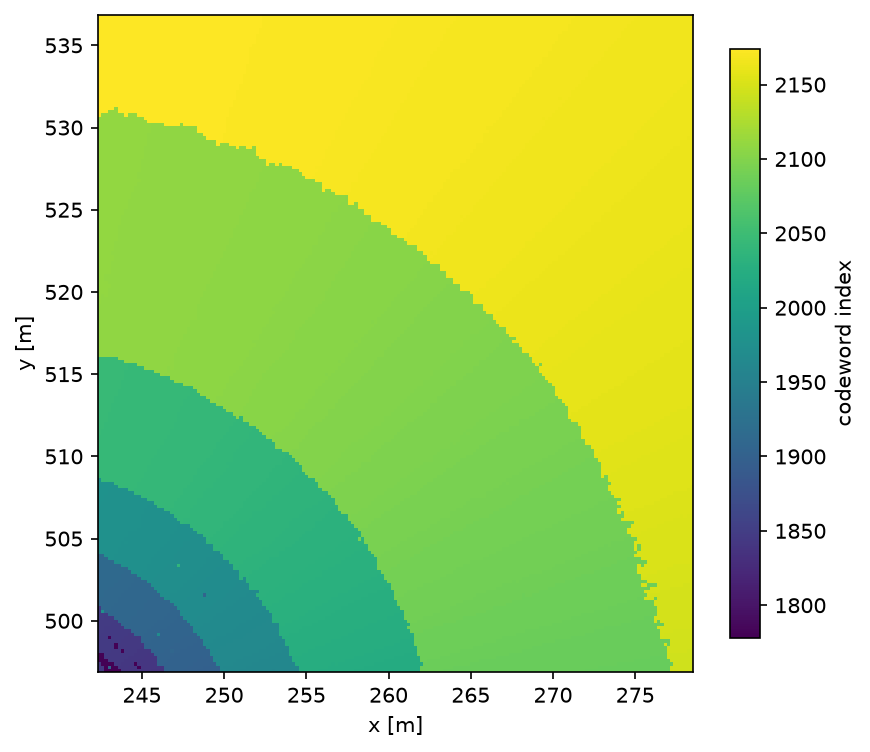}%
  \label{fig:beam_mlp}}\\
\subfloat[]{%
  \includegraphics[width=0.47\columnwidth]{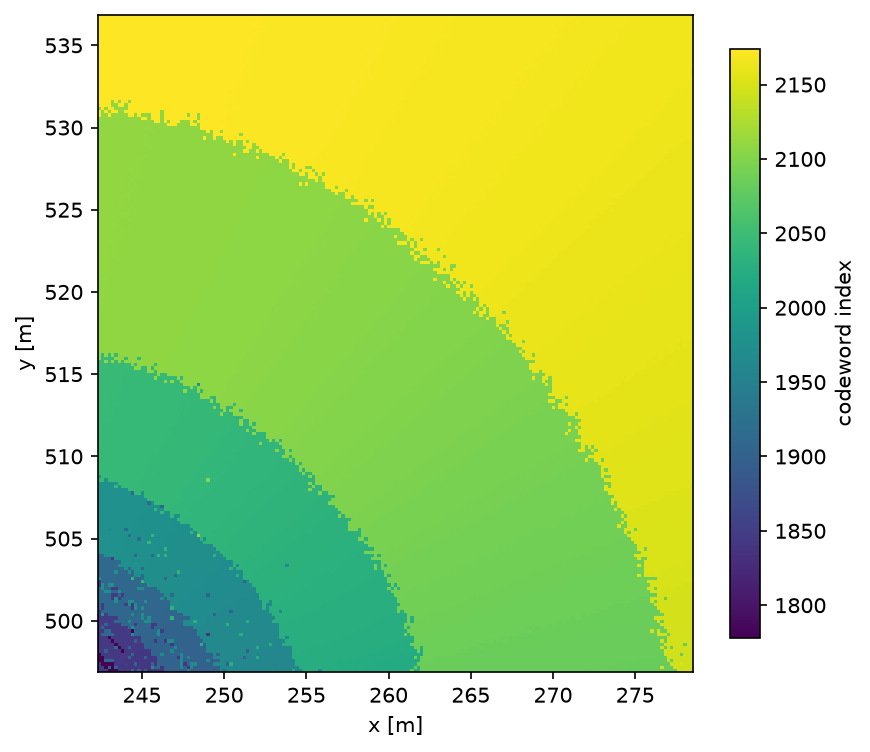}%
  \label{fig:beam_direct}}
\hfil
\subfloat[]{%
  \includegraphics[width=0.47\columnwidth]{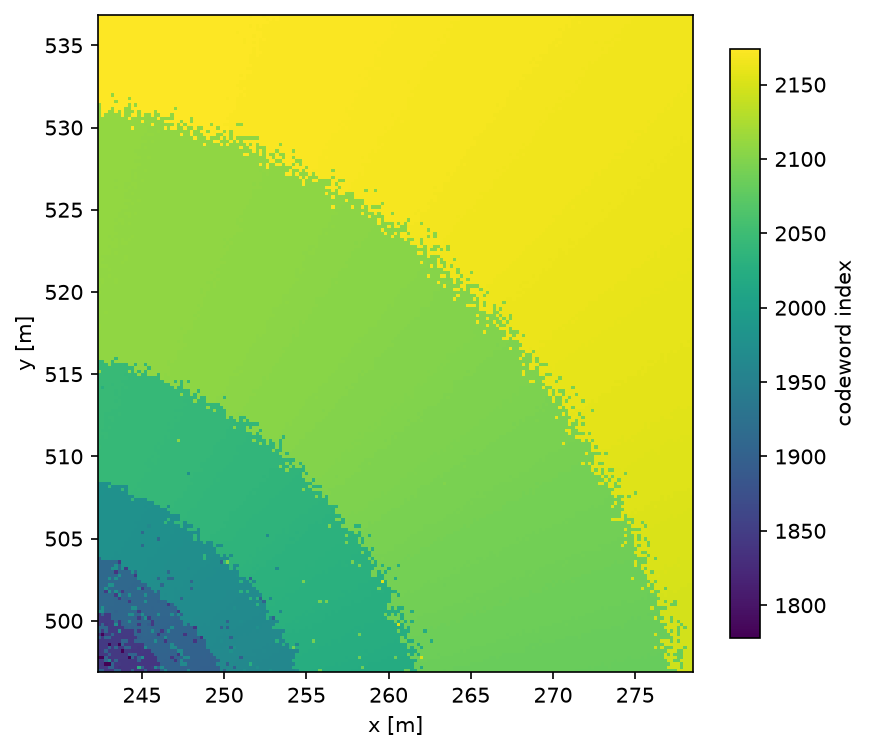}%
  \label{fig:beam_rate}}
\caption{Selected reflection beamforming index over the receiver grid for the $64\times64$ RIS with $M_a=8$ active elements. (a) Optimal beams, (b) MLP baseline, (c) direct-coded SNN with $T=5$, and (d) rate-coded SNN with $T=5$.}
\label{fig:beammaps}
\end{figure}

Two types of accuracy are used for evaluating our implementations, the top-1 beam accuracy, i.e., the fraction of evaluation users for which the selected codeword coincides with the optimal one, and the rate accuracy defined as the ratio between the rate achieved by the selected beam and the optimal rate. Both accuracies are evaluated on the first subcarrier, so that (\ref{eq:rate}) reduces to its $k=1$ term, and averaged over the $6200$ held-out users.

For the $64\times64$ RIS with $M_a=8$ and $30000$ training users, the MLP attains a top-1 beam accuracy of $95.2\%$ and a rate accuracy of $99.65\%$, the direct-coded SNN $90.5\%$ and $98.18\%$, and the rate-coded SNN $87.4\%$ and $96.92\%$. For every model, the rate accuracy exceeds the top-1 beam accuracy, i.e., most mis-selected codewords provide a gain close to the optimal one. This is consistent with Fig.~\ref{fig:beammaps}, where mis-selections appear at the boundaries between adjacent codewords, whose beams are nearly equivalent. On the training users, the top-1 beam accuracy is slightly higher, namely $95.7\%$, $91.7\%$ and $88.9\%$, indicating that all three models generalize to the interleaved evaluation positions.

\begin{figure}[t]
    \centering
    \includegraphics[width=\columnwidth]{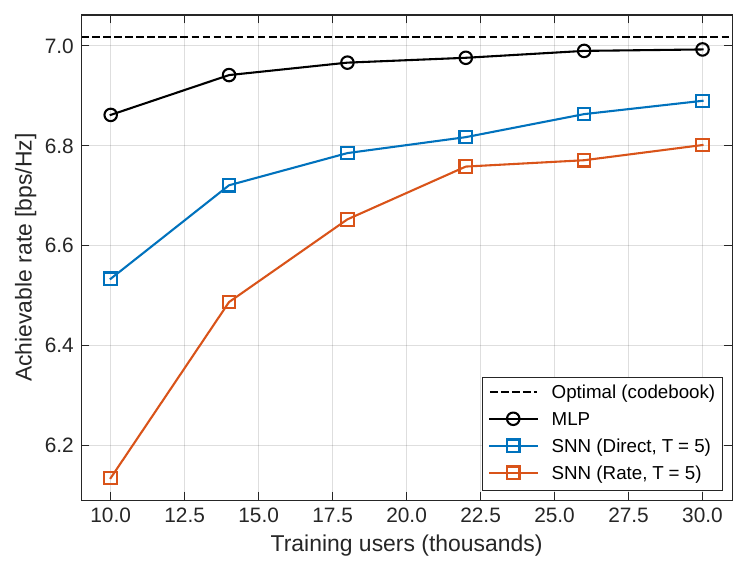}
    \caption{Achievable rate against the number of training users for $M_a=8$ and $T=5$.}
    \label{fig:rates}
\end{figure}

Fig.~\ref{fig:rates} shows the achievable rate against the number of training users, with the optimal rate of $7.02~\mathrm{bps/Hz}$ as reference. All models improve monotonically with the training set size, but the spiking models require more data: with $10000$ training users, the MLP attains a rate accuracy of $97.8\%$, against $93.1\%$ and $87.4\%$ for the direct- and rate-coded SNNs, and the gaps narrow at $30000$ users. Among the brain-inspired models, direct coding attains the higher accuracy, while the lower accuracy of rate coding is consistent with studies reporting an accuracy penalty for rate encoding \cite{Kim2022}.

\subsection{Energy Efficiency and Rate--Energy Trade-off}

\begin{figure}[!t]
    \centering
    \includegraphics[width=\columnwidth]{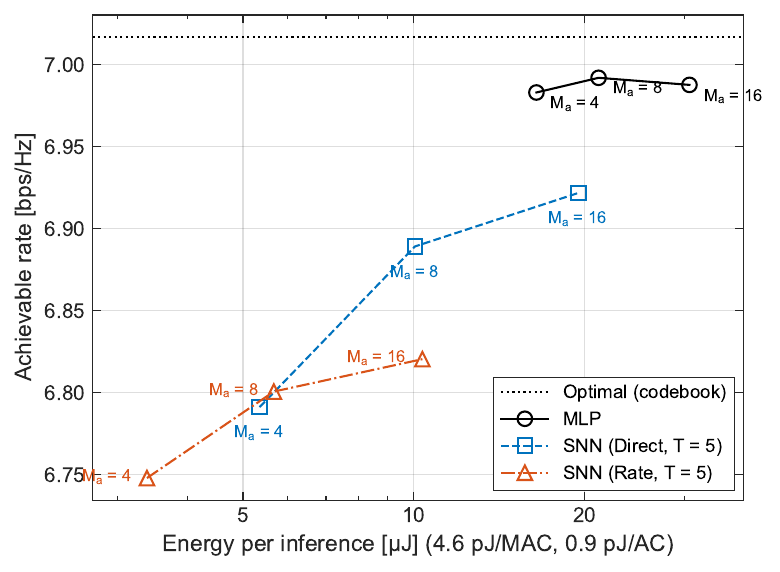}
    \caption{Achievable rate against inference energy for $M_a\in\{4,8,16\}$ active elements, for the $64\times64$ RIS with $T=5$ and $30000$ training users.}
    \label{fig:rate_energy}
\end{figure}

Next, the inference energy of the three models is compared, using (\ref{eq:consumption}) with $E_\mathrm{MAC}=4.6~\mathrm{pJ}$ and $E_\mathrm{AC}=0.9~\mathrm{pJ}$ for a $45~\mathrm{nm}$, $32$-bit design \cite{Horowitz2014}. For $M_a=8$, the MLP requires $21.14~\mathrm{\mu J}$ per inference, against $10.04~\mathrm{\mu J}$ for the direct-coded and $5.67~\mathrm{\mu J}$ for the rate-coded SNN, i.e., $2.1$ times and $3.7$ times less energy at a rate accuracy loss of $1.5$ and $2.7$ percentage points. The spiking layers are sparse, with the hidden neurons firing on average $0.10$ and $0.14$ times per time step, so that $\zeta^{(l)}\approx0.5$ and $0.7$, well below the break-even value $E_\mathrm{MAC}/E_\mathrm{AC}\approx5.1$. The consumption of both SNNs is therefore dominated by the input layer, which accounts for $93.8\%$ and $81.3\%$ of the total, against $44.6\%$ for the MLP. Under rate coding, the normalized features are symmetric around zero, so the sigmoid encoder fires at an average rate of $0.5$ per time step, i.e., $\zeta^{(1)}=T/2$ in (\ref{eq:rc_snn_flops}). Comparing the first term of (\ref{eq:rc_snn_flops}) with that of (\ref{eq:mac_mlp}) and (\ref{eq:dc_snn_flops}), the rate-coded input layer costs $TE_\mathrm{AC}/(2E_\mathrm{MAC})\approx0.49$ of its MAC counterpart at $T=5$. By the break-even condition $\zeta^{(1)}<E_\mathrm{MAC}/E_\mathrm{AC}$, this advantage vanishes for $T>2E_\mathrm{MAC}/E_\mathrm{AC}\approx10$, a regime not considered in Fig.~\ref{fig:rate_energy}.

Fig.~\ref{fig:rate_energy} shows the resulting rate--energy trade-off over the number of active elements. The MLP saturates already at $M_a=4$, with a rate accuracy of $99.5$--$99.7\%$, whereas both SNNs benefit from additional active elements, from $96.8\%$ to $98.7\%$ under direct coding and from $96.2\%$ to $97.2\%$ under rate coding. Since every input feature adds $500$ synaptic operations to the input layer, the energy of all models grows with $M_a$. The MLP and the direct-coded SNN grow identically, so the direct-coded SNN saves a constant $11.1~\mathrm{\mu J}$, and its energy reduction shrinks from $3.1$ times at $M_a=4$ to $1.6$ times at $M_a=16$. The rate-coded SNN grows at half the slope, retaining a reduction of $4.9$ times to $3.0$ times.

The non-dominated configurations are the rate-coded SNN, the direct-coded SNN and the MLP, each at $M_a\in\{4,8\}$. At $M_a=16$, the direct-coded SNN ($98.65\%$, $19.5~\mathrm{\mu J}$) is dominated by the MLP at $M_a=4$ ($99.52\%$, $16.4~\mathrm{\mu J}$), and the rate-coded SNN by the direct-coded SNN at $M_a=8$. As every active element also requires an RF chain, a small $M_a$ is preferable for the RIS, with the encoder selecting the operating point on the front, i.e., direct coding for a higher rate and rate coding for a lower energy.

\section{Conclusion}\label{section:5}

In this letter, an SNN-based beam selection scheme was presented for a semi-active RIS, which predicts the achievable rate of every codeword from channels sampled at a few active elements. For a $64\times64$ RIS, direct and rate coding attain $98.2\%$ and $96.9\%$ of the optimal rate at $2.1$ and $3.7$ times lower inference energy than an MLP of identical architecture, which attains $99.7\%$. As the input layer dominates the SNN energy, the advantage of direct coding shrinks as the number of active elements grows, whereas rate coding retains a $3.0$ to $4.9$ times reduction. The encoder thus selects the operating point on the rate--energy front.

\bibliographystyle{ieeetr}
\bibliography{IEEEabrv,References}

\end{document}